\documentclass{llncs}
\usepackage{fancyvrb}
\usepackage{amsmath}
\usepackage{amssymb}
\usepackage{amsfonts}
\usepackage{color}
\usepackage{colortbl}
\usepackage{fancybox}
\usepackage{url}
\usepackage{ifthen}
\usepackage[]{subfig}
\usepackage{tikz}
\usepackage{listings}
\usepackage{verbatim}
\usepackage{multicol}
\usepackage{listings}
\usepackage[utf8]{inputenc}
\usepackage{graphicx}
\usepackage[absolute]{textpos}

\lstnewenvironment{tabularlstlisting}[1][]
 {%
  \lstset{aboveskip=2mm,belowskip=-3.5ex,language=C,numbers=left,frame=none,escapechar=|,morekeywords={not},#1}%
 }
 {}

\lstnewenvironment{tabularlstlisting1}[1][]
 {%
  \lstset{aboveskip=-1.3ex,belowskip=-3.5ex,language=Ada,numbers=left,frame=none,escapechar=|,#1}%
 }
 {}

\newcounter{tbdcounter}
\newcommand\synchronized{synchronized}
\newcommand\Synchronized{Synchronized}
\newcommand\concurrent{concurrent}
\newcommand\Concurrent{Concurrent}

\title{Safe Non-blocking Synchronization in Ada~202x}
\author{%
Johann Blieberger\inst{1}
and
Bernd Burgstaller\inst{2}}
\institute{
Institute of Computer Engineering, Automation Systems Group,
TU Wien, Austria
\and
Department of Computer Science, Yonsei University, Korea
}

\begin{document}
\maketitle
\begin{abstract}
The mutual-exclusion property of locks stands in the way to scalability of
parallel programs on many-core architectures. Locks do not allow progress
guarantees, because a task may fail inside a critical section and keep holding
a lock that blocks other tasks from accessing shared data.  With non-blocking 
synchronization, the drawbacks of locks are avoided by synchronizing
access to shared data by atomic read-modify-write operations.  

To incorporate non-blocking synchronization in Ada~202x, programmers must be able to
reason about the behavior and performance of tasks in the absence of protected
objects and rendezvous.  We therefore extend Ada's memory model by synchronized types,
which support the
expression of memory ordering operations at a sufficient level of detail. To
mitigate the complexity associated with non-blocking synchronization, we propose
concurrent objects as a novel high-level language construct. Entities of a
concurrent object execute in parallel, due to a fine-grained, optimistic
synchronization mechanism. Synchronization is framed by the semantics of
concurrent entry execution. The programmer is only required to label shared
data accesses in the code of concurrent entries. Labels constitute
memory-ordering operations expressed through attributes.  To the best of our
knowledge, this is the first approach to provide a non-blocking synchronization 
construct as a first-class citizen of a high-level programming language. We
illustrate the use of concurrent objects
by several examples.
\end{abstract}

\begin{section}{Introduction\label{sec:intro}}

Mutual exclusion locks are the most common technique to synchronize multiple
tasks to access shared data. Ada's protected objects (POs) implement the
monitor-lock concept~\cite{Hoare:monitors}. Method-level locking requires a
task to acquire an exclusive lock to execute a PO's entry or procedure.
(Protected functions allow concurrent read-access in the style
of a readers--writers lock~\cite{Herlihy:2012}.) Entries
and procedures of a PO thus effectively execute one after another, which makes
it straight-forward for programmers to reason about updates to the shared data
encapsulated by a PO.
Informally, sequential consistency ensures that method calls act as if they
occurred in a sequential, total order that is consistent with the program order
of each participating task. I.e., for any concurrent execution, the method calls
to POs can be ordered sequentially such that they (1)~are consistent with program
order, and (2)~meet each PO's specification (pre-condition, side-effect,
post-condition)~\cite{Herlihy:2012}.

Although the sequential consistency semantics of mutual exclusion locks
facilitate reasoning about programs, they nevertheless introduce potential
concurrency bugs such as dead-lock, live-lock and priority inversion. The
mutual-exclusion property of (highly-contended) locks stands in the way to
scalability of parallel programs on many-core architectures~\cite{Scott:Sync}.
Locks do not allow progress guarantees, because a task may fail inside a
critical section (e.g., by entering an endless loop), preventing other tasks
from accessing shared data. 

Given the disadvantages of mutual exclusion locks, it is thus desirable to
give up on method-level locking and allow method calls
to overlap in time. Synchronization is then performed on a finer granularity
within a method's code, via atomic {\em read-modify-write}~(RMW) operations. In
the absence of mutual exclusion locks, the possibility of task-failure inside a
critical section is eliminated, because critical sections are reduced to single
atomic operations. These atomic operations are provided either by the CPU's
instruction set architecture (ISA), or the language run-time (with the help of
the CPU's ISA). It thus becomes possible to provide progress guarantees, which
are unattainable with locks. In particular, a method is {\em non-blocking\/},
if a task's pending invocation is never required to wait for another task's
pending invocation to complete~\cite{Herlihy:2012}.

\begin{figure}[b]
\begin{center}
\begin{tabular}{ll@{\hskip 11mm}l}
{\begin{lstlisting}[language=C, firstnumber=1, frame=none, escapechar=|]
-- Initial values:
Flag := False;
Data := 0;
\end{lstlisting}}\\[4mm]
{\begin{tabularlstlisting}
-- Task 1:
Data := 1;     |\label{l:t1:store:1}|
Flag := True;  |\label{l:t1:store:2}|
\end{tabularlstlisting}}
&
{
\begin{tabularlstlisting}
-- Task 2:
loop|\label{l:spin:1}|
  R1 := Flag;
  exit when R1;
end loop; |\label{l:spin:2}|
R2 := Data; |\label{l:read}|
\end{tabularlstlisting}
}
&
{
\begin{tabularlstlisting1}
Data : Integer with Volatile;
Flag : Boolean with Atomic;
\end{tabularlstlisting1}
}\\[5mm]
\multicolumn{2}{c}{(a)\rule{18mm}{0mm}}&\multicolumn{1}{c}{(b)} 
\end{tabular}
\end{center}
\vspace{-4mm}
\caption{(a) Producer-consumer synchronization in pseudo-code: Task~1 writes the {\tt Data} variable and
         then signals Task~2 by setting the {\tt Flag} variable.
         Task~2 is spinning on the {\tt Flag} variable (lines~\ref{l:spin:1} to~\ref{l:spin:2}) and then reads the
         {\tt Data} variable. (b) Labeling to enforce sequential consistency in Ada~2012.} 
\label{fig:mem:example}
\end{figure}

Non-blocking synchronization techniques are notoriously difficult to implement
and the design of non-blocking data structures is an area of active research.
To enable non-blocking synchronization, a programming language must provide a
strict memory model. The purpose of a memory model is to define the set of
values a read operation in a program is allowed to
return~\cite{Boehm:mem:survey}.

To motivate the need for a strict memory model, consider the producer-consumer
synchronization example in Fig.~\ref{fig:mem:example}(a) (adopted
from~\cite{Hill:CSprimer} and~\cite{Barnes:RationaleAda2012}).  The
programmer's intention is to communicate the value of variable~{\tt Data} from
Task~1 to Task~2. Without explicitly requesting a sequentially consistent
execution, a compiler or CPU may break the programmer's intended
synchronization via the {\tt Flag} variable by re-ordering memory operations
that will result in reading {\tt R2 = 0} in Line~\ref{l:read} of Task~2. (E.g.,
a store--store re-ordering of the assignments in lines~\ref{l:t1:store:1}
and~\ref{l:t1:store:2} of Task~1 will allow this result.) In Ada~2012, such
re-orderings can be ruled out by labeling variables {\tt Data} and {\tt Flag}
by aspect {\tt volatile}.  The corresponding variable declarations are depicted
in Fig.~\ref{fig:mem:example}(b).  (Note that by \cite[C.6§8/3]{Ada2012LRM}
aspect {\tt atomic} implies aspect {\tt volatile}, but not vice versa.) 

The intention for volatile variables in Ada~2012 was to guarantee that all
tasks agree on the same order of updates~\cite[C.6§16/3]{Ada2012LRM}.  Updates
of volatile variables are thus required to be sequentially consistent, in the
sense of Lamport's definition~\cite{Lamport:SC}: ``{\em With sequential
consistency (SC), any execution has a total order over all memory writes and
reads, with each read reading from the most recent write to the same
location}''.

However, the Ada~2012 aspect~{\tt volatile} has the following shortcomings:
\begin{enumerate}
\item 
Ensuring SC for multiple tasks without atomic access is impossible.
Non-atomic volatile variables therefore should not be provided by the language.
Otherwise, the responsibility shifts from the programming language
implementation to the programmer to ensure SC
by pairing an atomic (implied volatile) variable with each non-atomic
volatile variable (see, e.g., Fig.~\ref{fig:mem:example}(b) and
\cite{Simpson1990} for examples).
(Note that a programming language implementation may ensure
atomicity by a mutual exclusion lock if
no hardware primitives for atomic access to a particular type of shared data are available.) 
\item
Requiring SC on all shared variables is costly in terms of performance on
contemporary multi-core CPUs.  In Fig.~\ref{fig:mem:example}, performance can
be improved by allowing a less strict memory order for variable~{\tt Data} (to
be addressed in Section~\ref{sec:mm}).
\item
Although Ada provides the highly abstract PO monitor-construct
for blocking synchronization, there is currently no programming primitive
available to match this abstraction level for non-blocking synchronization. 
\end{enumerate}

Contemporary CPU architectures relax SC for the sake of
performance~\cite{Adve:SMC:tutorial,Hennessy:acquire:release,Hill:CSprimer}.
It is a challenge for programming language designers to provide safe, efficient
and user-friendly non-blocking synchronization features.  The original memory
model for Java contained problems and had to be revised~\cite{Manson:JMM}.  It was
later found to be unsound with standard compiler
optimizations~\cite{Sevcik:JVM:CC:problems}.  The C++11 standard (cf.\
\cite{c++11,will:c++}) has already specified a strict memory model for
concurrent and parallel computing.  We think that C++11 was not entirely
successful both in terms of safety and in terms of being user-friendly.
In contrast, we are convinced that these challenges can be met in the upcoming
Ada~202x standard.

It has been felt since Ada~95 that it might be advantageous to have language
support for synchronization based on atomic variables.  For example, we
cite~\cite[C.1]{Ada95rat}: {\it ``A need to access specific machine
instructions arises sometimes from other considerations as well. Examples
include instructions that perform compound operations atomically on shared
memory, such as test-and-set and compare-and-swap, and instructions that
provide high-level operations, such as translate-and-test and vector
arithmetic.''}

Ada is already well-positioned to provide a strict memory model in conjunction with
support for non-blocking synchronization, because it provides tasks 
as first-class citizens. This rules out inconsistencies that may result from
thread-functionality provided through libraries~\cite{Boehm:thread:library}.

To provide safe and efficient non-blocking synchronization for
Ada~202x, this paper makes the following contributions:

\begin{enumerate}
\item
We extend Ada's memory model by introducing synchronized types,
which allow the expression of memory ordering operations consistently and
at a sufficient level of detail. Memory ordering operations are expressed through aspects and attributes.
Language support for spin loop synchronization via \synchronized\ variables is proposed.
\item
We propose {\em concurrent objects\/} (COs) as a high-level language
construct to express non-blocking synchronization. COs are meant to encapsulate
the intricacies of non-blocking synchronization as POs do for blocking synchronization.
Contrary to POs, the entries and procedures of COs execute in parallel, due to 
a fine-grained, optimistic
synchronization mechanism.
\item
We provide an alternative, low-level API on synchronized types, which provides
programmers with full control over the implementation of non-blocking synchronization
semantics. Our main purpose with the low-level API is to provoke a discussion on the
trade-off between abstraction versus flexibility.
\item
We illustrate the use of concurrent objects and the alternative, low-level API
by several examples.
\end{enumerate}

The remainder of this paper is organized as follows.  We summarize the
state-of-the-art on memory models and introduce synchronized variables in
Sec.~\ref{sec:mm}.  We introduce attributes for specifying memory ordering
operations in Sec.~\ref{aspects}.  We specify concurrent objects in
Sec.~\ref{nbPO} and discuss task scheduling in the presence of COs in
Sec.~\ref{sched}.  Sec.~\ref{sec:examples} contains two CO example implementations with
varying memory consistency semantics.  We discuss our low-level API in
Sec.~\ref{API_sec}.
Sec.~\ref{sec:concl} contains our conclusions.

This paper is an extension of work that
appeared at the Ada-Europe~2018 conference~\cite{BB:2018}.
Additional material is confined to two appendices: Appendix~\ref{sec:rationale}
states the design-decisions of our proposed non-blocking synchronization mechanisms.
Appendix~\ref{app:examples} contains further examples.
%
\end{section}

\begin{section}{The Memory Model\label{sec:mm}}

For reasons outlined in Sec.~\ref{sec:intro}, we do not consider the Ada~2012
atomic and volatile types here. Rather, we introduce {\em\synchronized} types
and variables.  Synchronized types provide atomic access. We propose aspects and attributes
for specifying a particular memory model to be employed for reading/writing
\synchronized\ variables.


Modern multi-core computer architectures are equipped with a memory hierarchy
that consist of main memory, caches and registers.
It is important to distinguish between memory
consistency and coherence.  We cite from~\cite{Hill:CSprimer}: {\it `For a
shared memory machine, the memory consistency model defines the architecturally
visible behavior of its memory system. Consistency definitions provide rules
about loads and stores (or memory reads and writes) and how they act upon
memory. As part of supporting a memory consistency model, many machines also
provide cache coherence protocols that ensure that multiple cached copies of
data are kept up-to-date.'}

The purpose of a memory consistency model (or memory model, for short) is to
define the set of values a read operation is allowed to
return~\cite{Boehm:mem:survey}. To facilitate programmers' intuition, it would
be ideal if all read/write operations of a program's tasks are sequentially
consistent. However, the hardware memory models provided by contemporary CPU
architectures relax SC for the sake of
performance~\cite{Adve:SMC:tutorial,Hennessy:acquire:release,Hill:CSprimer}.
Enforcing SC on such architectures may incur a noticeable performance penalty.
The workable middle-ground between intuition (SC) and performance (relaxed
hardware memory models) has been established with SC for data race-free
programs (SC-for-DRF)~\cite{Hill:DRF:SC}.  Informally, a program has a data
race if two tasks access the same memory location, at least one of them is a
write, and there are no intervening synchronization operations that would order
the accesses.  ``SC-for-DRF'' requires programmers to ensure that programs are
free of data races under SC. In turn, the relaxed memory model of a SC-for-DRF
CPU guarantees SC for all executions of such a program.

It has been acknowledged in the
literature~\cite{Boehm:mem:survey}  that Ada~83 was perhaps the first
widely-use high-level programming language to provide first-class support for
shared-memory programming.  The approach taken with Ada~83 and later language
revisions was to require legal
programs to be without synchronization errors, which is the approach taken with
SC-for-DRF. In contrast, for the Java memory model it was perceived that even
programs with synchronization errors shall have defined semantics for reasons
of safety and security of Java's sand-boxed execution environment. (We do not
consider this approach in the remainder of this paper, because it does not
align with Ada's current approach to regard the semantics of programs with
synchronization errors as undefined, i.e., as an {\em erroneous execution},
by~\cite[9.10§11]{Ada2012LRM}.) The SC-for-DRF programming model and two
relaxations were formalized for C++11~\cite{Boehm:cpp:mem}.  They were later
adopted for C11, OpenCL~2.0, and for X10~\cite{Zwinkau:x10:mem} (without the
relaxations).

On the programming language level to guarantee DRF, means for synchronization
(ordering operations) have to be provided.  Ada's POs are well-suited for this
purpose. For non-blocking synchronization, atomic operations can be used to
enforce an ordering between the memory accesses of two tasks.
It is one goal of this paper to add language features to Ada such that atomic operations 
can be employed with DRF programs.  To avoid ambiguity, we propose
\synchronized\ variables and types, which support the expression of memory
ordering operations at a sufficient level of detail (see
Sec.~\ref{sec:sync:var}).

The purpose of \synchronized\ variables is that they can be used to safely
transfer information (i.e., the value of the variables) from one task to
another.  ISAs provide atomic load/store instructions only for a limited set of
primitive types.  Beyond those, atomicity can only be ensured by locks.
Nevertheless, computer architectures provide memory fences (see e.g.,
\cite{Herlihy:2012}) to provide means for ordering memory operations. A memory
fence requires that all memory operations before the fence (in program order)
must be committed to the memory hierarchy before any operation after the fence.  Then, for data to be
transferred from one thread to another it is not necessary to be atomic
anymore. I.e., it is sufficient that (1)~the signaling variable is atomic, and
that (2)~all write operations are committed to the memory hierarchy before setting the signaling
variable. On the receiver's side, it must be ensured that (3)~the signaling
variable is read atomically, and that (4)~memory loads for the data occur
after reading the signaling variable (Listing~\ref{genRA} provides an example.)


In addition to \synchronized\ variables, \synchronized\ types and attribute\linebreak[4]
{\tt \Synchronized\_Components} are convenient means for
enhancing the usefulness of \synchronized\ variables.

The general idea of our proposed approach is to define {\em non-blocking\/}
\concurrent\ objects similar to protected objects (cf.\ e.g.,
\cite{Herlihy:2012}).  However, entries of \concurrent\ objects will not block
on guards; they will spin loop until the guard evaluates to true.  In addition,
functions, procedures, and entries of \concurrent\ objects are allowed to
execute and to modify the encapsulated data in parallel.  Private entries for
\concurrent\ objects are also supported.
It is their responsibility that the
data provides a consistent view to the users of the \concurrent\ object.
\Concurrent\ objects will use \synchronized\ types for synchronizing data
access. Several memory models are provided for doing this efficiently.  It is
the responsibility of the programmer to ensure that the entries of a
\concurrent\ object are free from data races (DRF). For such programs, the
non-blocking semantics of a \concurrent\ object will provide SC in the same way
as protected objects do for blocking synchronization.


\begin{subsection}{Synchronizing memory operations and enforcing ordering}

For defining ordering relations on memory operations, it is useful to introduce some other useful relations.

The {\em synchronizes-with\/} relation can be achieved only by use of atomic types.
Even if monitors or protected objects are used for synchronization, the runtime implements them employing atomic types.
The general idea is to equip read and write operations on an atomic variable with information that
will enforce an ordering on the read and write operations.
Our proposal is to use attributes for specifying this ordering information.
Details can be found below.

The {\em happens-before\/} relation is the basic relation for ordering operations in programs.
In a program consisting of only one single thread, happens-before is straightforward.
For inter-thread happens-before relations the synchronizes-with relation becomes important.
If operation {\tt X} in one thread synchronizes-with operation {\tt Y} in another thread,
then {\tt X} happens-before {\tt Y}. Note that the happens-before relation is transitive,
i.e., if {\tt X} happens-before {\tt Y} and {\tt Y} happens-before {\tt Z}, then {\tt X} happens-before {\tt Z}.
This is true even if {\tt X}, {\tt Y}, and {\tt Z} are part of different threads.

We define different memory models.  These memory models originated from the
DRF~\cite{Hill:DRF:SC} and properly-labeled~\cite{Hennessy:acquire:release}
hardware memory models. They were formalized for the memory model of
C++~\cite{Boehm:cpp:mem}.  The ``sequentially consistent'' and
``acquire-release'' memory models provide SC for DRF.  The models can have
varying costs on different computer architectures.  The ``acquire-release''
memory model is a relaxation of the ``sequentially consistent'' memory model.
As described in Table~\ref{MMcompcpu}, it requires concessions from the
programmer to weaken SC in turn for more flexibility for 
the CPU to re-order memory operations.

\begin{description}
\item[Sequentially Consistent Ordering] is the most stringent model and the easiest one
for programmers to work with.
In this case all threads see the same, total order of operations.
This means, a sequentially consistent write to a \synchronized\ variable synchronizes-with a sequentially-consistent read of the same variable.
\item[Relaxed Ordering] does not obey synchronizes-with relationships, but operations on the same \synchronized\ variable within a single thread
still obey happens-before relationships.
This means that although one thread may write a \synchronized\ variable, at a later point in time another thread may read 
an earlier value of this variable.
\item[Acquire-Release Ordering] when compared to relaxed ordering introduces some synchronization.
In fact, a read operation on \synchronized\ variables can then be labeled by {\em acquire},
a write operation can be labeled by {\em release}.
Synchronization between release and acquire is pairwise between the thread that issues the release and that acquire operation of a thread that does the first read-acquire after the release.\footnote{In global time!}
A thread issuing a read-acquire later may read a different value than that written by the first thread.

\end{description}
\end{subsection}

\begin{table}[t]
\begin{tabular}{p{25mm}|c|p{80mm}}
{\bf memory order} & {\bf involved} & {\bf constraints for reordering memory accesses}\\
& {\bf threads} & {\bf (for compilers and CPUs)}\\
\hline
relaxed & 1 & no inter-thread constraints \\
\hline
release/acquire& 2 &
(1) ordinary\footnotemark\ stores originally\footnotemark\ before {\tt release} (in program order) will happen before the release fence (after compiler optimizations and CPU reordering)\newline
(2) ordinary loads originally after {\tt acquire} (in program order) will take place after the acquire fence (after compiler optimizations and CPU reordering)\\
\hline
sequentially\newline consistent & all &
(1) all memory accesses originally before the {\tt sequenti\-ally\_consistent} one (in program order) will happen before the fence (after compiler optimizations and CPU reordering)\newline
(2) all memory accesses originally after the {\tt sequenti\-ally\_consistent} one (in program order) will happen after the fence (after compiler optimizations and CPU reordering)
\end{tabular}\\\ \\
\caption{Memory Order and Constraints for Compilers and CPUs\vspace*{-2\baselineskip}}\label{MMcompcpu}
\end{table}
\addtocounter{footnote}{-1}
\footnotetext{Memory accesses other than accesses to \synchronized\ variables}
\addtocounter{footnote}{1}
\footnotetext{Before optimizations performed by the compiler and before reordering done by the CPU.}

It is important to note, that the semantics of the models above have to be
enforced by the compiler (for programs which are DRF).  I.e., the compiler ``knows'' the relaxed memory
model of the hardware and inserts memory fences in the machine-code such that
the memory model of the high-level programming language is enforced.
Compiler optimizations must ensure that
reordering of operations is performed in such a way that the semantics of
the memory model are not violated.  The same applies to CPUs, i.e., reordering
of instructions is done with respect to the CPU's relaxed hardware memory model,
constrained by the ordering semantics of fences inserted by the compiler.  The constraints
enforced by the memory model are summarized in Table~\ref{MMcompcpu}.
\end{section}

\begin{section}{Synchronization primitives}\label{aspects}

%

\begin{subsection}{\Synchronized\ Variables\label{sec:sync:var}}
\Synchronized\ variables can be used as atomic variables in Ada~2012, the only exception being that they are declared inside the lexical scope (data part) of a \concurrent\ object.
In this case aspects and attributes used in the implementation of the \concurrent\ object's operations (functions, procedures, and entries) are
employed for specifying behavior according to the memory model.
Variables are labeled by the boolean aspect {\tt \Synchronized}.

Read accesses to \synchronized\ variables in the implementation of the \concurrent\ object's operations may be labeled with the attribute {\tt Concurrent\_Read},
write accesses with the attribute {\tt Concurrent\_Write}.
Both attributes have a parameter {\tt Memory\_Order} to specify the memory order of the access.
(If the operations are not labeled, the default values given below apply.)
In case of read accesses, values allowed for parameter {\tt Memory\_Order} are
{\tt Sequentially\_Consistent},
{\tt Acquire}, and
{\tt Relaxed}.
The default value is
{\tt Sequentially\_Consistent}.
For write accesses the values allowed are
{\tt Sequentially\_Consistent},
{\tt Release}, and
{\tt Relaxed}.  
The default value is again {\tt Sequentially\_Consistent}.

For example, assigning the value of \synchronized\ variable {\tt Y} to \synchronized\ variable {\tt X} is given like\newline
\hspace*{5mm}{\tt X'Concurrent\_Write(Memory\_Order => Release) :=\newline\hspace*{30mm}\hfill Y'Concurrent\_Read(Memory\_Order => Acquire);}

%
In addition we propose aspects for specifying variable specific default values for the attributes described above.
In more detail, when declaring \synchronized\ variables the default values for read and write accesses can be specified via aspects
{\tt Memory\_Order\_Read} and {\tt Memory\_Order\_Write}.
The allowed values are the same as those given above for read and write accesses.
If these memory model aspects are given when declaring a \synchronized\ variable,
the attributes {\tt Concurrent\_Read} and {\tt Concurrent\_Write} need not be given for actual read and write accesses of this variable.
However, these attributes may be used to temporarily over-write the default values specified for the variable by the aspects.
For example\\
\hspace*{5mm}{\tt X: integer with Synchronized, Memory\_Order\_Write => Release;}\\
\hspace*{5mm}{\tt Y: integer with Synchronized, Memory\_Order\_Read => Acquire;}\\
\dots\\
\hspace*{5mm}{\tt X := Y};\\
does the same as the example above but without spoiling the assignment statement.

Aspect {\tt Synchronized\_Components} relates to aspect {\tt Synchronized} in the same way as
{\tt Atomic\_Components} relates to {\tt Atomic} in Ada~2012.

\end{subsection}

\begin{subsection}{Read-Modify-Write Variables}
If a variable inside the data part of a \concurrent\ object is labeled by the
aspect {\tt Read\_Modify\_Write}, this implies that the variable is \synchronized.
Write access to a read-modify-write variable in the implementation of the protected object's operations
is a read-modify-write access.
The read-modify-write access is done via the attribute {\tt Concurrent\_Exchange}.
The two parameters of this attribute are {\tt Memory\_Order\_Success} and {\tt Memory\_Order\_Failure}.
The first specifies the memory order for a successful write, the second one the memory order if the write access fails (and a new value is assigned to the variable).

{\tt Memory\_Order\_Success} is one of 
{\tt Sequentially\_Consistent},
{\tt Acquire},\linebreak[4]
{\tt Release}, and
{\tt Relaxed}.  

{\tt Memory\_Order\_Failure} may be one of
{\tt Sequentially\_Consistent},
{\tt Acquire}, and
{\tt Relaxed}.  
The default value for both is {\tt Sequentially\_Consistent}.
For the same read-modify-write access the memory order specified for failure must not be stricter than that specified for success.
So, if {\tt Memory\_Order\_Failure => Acquire} or {\tt Memory\_Order\_Failure => Sequentially\_Consistent} is specified,
these have also be given for success.

For read access to a read-modify-write variable, attribute {\tt Concurrent\_Read} has to be used.
The parameter {\tt Memory\_Order} has to be given.
Its value is one of
{\tt Sequentially\_Consistent},
{\tt Acquire},
{\tt Relaxed}.
The default value is\linebreak[4]
{\tt Sequentially\_Consistent}.

Again, aspects for variable specific default values for the attributes described above may be specified when declaring a read-modify-write variable.
The aspects are {\tt Memory\_Order\_Read}, {\tt Memory\_Order\_Write\_Success}, and\linebreak[4]
{\tt Memory\_Order\_Write\_Failure} with allowed values as above.
\end{subsection}

\begin{subsection}{Synchronization Loops}
As presented below synchronization by \synchronized\ variables is performed via spin loops.
We call these loops {\em sync loops}.
\end{subsection}

\end{section}

\begin{section}{\Concurrent\ Objects}\label{nbPO}

\begin{subsection}{Non-Blocking Synchronization}

Besides the aspects and attributes proposed in Section~\ref{aspects} that have to be used for implementing \concurrent\ objects,
\concurrent\ objects are different from protected objects in the following way.
All operations of \concurrent\ objects can be executed in parallel.
\Synchronized\ variables have to be used for synchronizing the executing operations.
Entries have Boolean-valued guards.
The Boolean expressions for such guards may contain only \synchronized\ variables declared in the data part of the protected object
and constants.
Calling an entry results either in immediate execution of the entry's body if the guard evaluates to {\tt true}, or
in spin-looping until eventually the guard evaluates to {\tt true}.
We call such a spin loop {\em sync loop}.

\end{subsection}

\begin{subsection}{Read-Modify-Write Synchronization}

For \concurrent\ objects with read-modify-write variables the attributes proposed in Section~\ref{aspects} apply.
All operations of \concurrent\ objects can be executed in parallel.
Read-modify-write variables have to be used for synchronizing the executing operations.
The guards of entries have to be of the form {\tt X = X'OLD} where {\tt X} denotes a read-modify-write variable of the
\concurrent\ object. The attribute {\tt OLD} is well-known from postconditions.
An example in our context can be found in Listing~\ref{lfsada}.

If during the execution of an entry a read-modify-write operation is reached, that operation might succeed immediately, in which case
execution proceeds after the operation in the normal way.
If the operation fails, the whole execution of the entry is restarted ({\em implicit sync loop\/}).
In particular, only the statements being data-dependent on the read-modify-write variable are re-executed.
Statements not being data-dependent on the read-modify-write variables are executed only on the first try.\footnote{
For the case that the compiler cannot figure out which statements are data-dependent, we propose an additional Boolean aspect
{\tt only\_execute\_on\_first\_try} to tag non-data-dependent statements.}
Precluding non-data-dependent statements from re-execution is not only a matter of efficiency, it sometimes makes sense semantically,
e.g., for adding heap management to an implementation.

\end{subsection}

\end{section}

\begin{section}{Scheduling and Dispatching}\label{sched}

We propose a new state for Ada tasks to facilitate correct scheduling and dispatching for threads synchronizing via \synchronized\ or read-modify-write types.
If a thread is in a sync loop, the thread state changes to ``in\_sync\_loop''.
Note that sync loops can only happen inside \concurrent\ objects.
Thus they can be spotted easily by the compiler and cannot be confused with ``normal'' loops.
Note also that for the state change it makes sense not to take place during the first iteration of the sync loop,
because the synchronization may succeed immediately.
For read-modify-write loops, iteration from the third iteration on may be a good choice;
for spin loops, an iteration from the second iteration on may be a good choice.

In this way the runtime can guarantee that not all available CPUs (cores) are occupied by threads in state ``in\_sync\_loop''.
Thus we can be sure that at least one thread makes progress and finally all \synchronized\ or read-modify-write variables are released (if the program's synchronization structure is correct and the program does not deadlock).

After leaving a sync loop, the thread state changes back to ``runable''.


\end{section}

\begin{section}{Examples}\label{sec:examples}

\begin{subsubsection}{Non-blocking Stack.}\label{lfs}
Listing~\ref{lfsada} shows an implementation of a non-blocking stack using our proposed new syntax for \concurrent\ objects.
%
\begin{lstlisting}[escapechar=|,
caption={Non-blocking Stack Implementation Using Proposed New Syntax},
label={lfsada}
]
  subtype Data is Integer;

  type List;
  type List_P is access List;
  type List is
    record
      D: Data;
      Next: List_P;
    end record;

  Empty: exception;

  concurrent Lock_Free_Stack
  is
    entry Push(D: Data);
    entry Pop(D: out Data);
  private
    Head: List_P with Read_Modify_Write,|\label{lfs:rmw}|
      Memory_Order_Read => Relaxed,|\label{lfs:mmb}|
      Memory_Order_Write_Success => Release,
      Memory_Order_Write_Failure => Relaxed;|\label{lfs:mme}|
  end Lock_Free_Stack;

  concurrent body Lock_Free_Stack is
    entry Push (D: Data)|\label{lfs:pushb}|
        until Head = Head'OLD is
      New_Node: List_P := new List;|\label{lfs:new}|
    begin
      New_Node.all := (D => D, Next => Head);|\label{lfs:newb}|
      Head := New_Node;|\label{lfs:rmwo}|
    end Push;|\label{lfs:pushe}|

    entry Pop(D: out Data)|\label{lfs:popb}|
        until Head = Head'OLD is
      Old_Head: List_P;
    begin
      Old_Head := Head;
      if Old_Head /= null then
        Head := Old_Head.Next;
        D := Old_head.D;
      else
        raise Empty;
      end if;
    end Pop;|\label{lfs:pope}|
  end Lock_Free_Stack;
\end{lstlisting}

Implementation of entry {\tt Push} (lines~\ref{lfs:pushb}--\ref{lfs:pushe}) behaves as follows.
In Line~\ref{lfs:newb} a new element is inserted at the head of the list.
Pointer {\tt Next} of this element is set to the current head.
The next statement (Line~\ref{lfs:rmwo}) assigns the new value to the head of the list.
Since variable {\tt Head} has aspect {\tt Read\_Modify\_Write} (line~\ref{lfs:rmw}), this is done with RMW semantics,
i.e., if the value of {\tt Head} has not been changed (since the execution of {\tt Push} has started) by a different thread executing {\tt Push} or {\tt Pop} (i.e., {\tt Head = Head'OLD}),
then the RMW operation succeeds and execution proceeds at Line~\ref{lfs:pushe}, i.e., {\tt Push} returns.
If the value of {\tt Head} has been changed ({\tt Head /= Head'OLD}),
then the RMW operation fails and entry {\tt Push} is re-executed starting from Line~\ref{lfs:newb}.
Line~\ref{lfs:new} is not re-executed as it is not data dependent on {\tt Head}.

Several memory order attributes apply to the RMW operation (Line~\ref{lfs:rmwo}) which are given in lines~\ref{lfs:mmb}--\ref{lfs:mme}:
In case of a successful execution of the RMW, the value of {\tt Head} is released such that other threads can read its value via memory order acquire.
In the failure case the new value of {\tt Head} is assigned to the ``local copy'' of {\tt Head} (i.e., {\tt Head'OLD}) via relaxed memory order.
``Relaxed'' is enough because RMW semantics will detect if the value of {\tt Head} has been changed by a different thread anyway.
The same applies to ``Relaxed'' in Line~\ref{lfs:new}.

Implementation of entry {\tt Pop} (lines~\ref{lfs:popb}--\ref{lfs:pope}) follows along the same lines.

Memory management needs special consideration:
In our case it is enough to use a \synchronized\ counter that counts the number of threads inside {\tt Pop}.
If the counter equals~$1$, memory can be freed.
Ada's storage pools are a perfect means for doing this without spoiling the code.

This example also shows how easy it is to migrate from a (working) blocking to a (working) non-blocking implementation of a program.
Assume that a working implementation with a protected object exists, then one has to follow these steps:
\begin{enumerate}
\item
Replace keyword {\tt protected} by keyword {\tt concurrent}.
\item
Replace protected operations by DRF \concurrent\ operations, thereby adding appropriate guards to the \concurrent\ entries.
\item
Test the non-blocking program which now has default memory order\linebreak[4]
{\tt sequentially\_consistent}.
\item
Carefully relax the memory ordering requirements: Add memory order aspects and/or attributes {\tt Acquire}, {\tt Release}, and/or {\tt Relaxed} to improve performance but without violating memory consistency.
\end{enumerate}
\end{subsubsection}

\begin{subsubsection}{Generic Release-Acquire Object.}
\enlargethispage{\baselineskip}
Listing~\ref{genRA} shows how release-acquire semantics can be implemented for general data structures with help of one synchronized Boolean.
\begin{lstlisting}[escapechar=|,
caption={Generic Release-Acquire Object},
label={genRA}
]
generic
  type Data is private;
package Generic_Release_Acquire is

  concurrent RA
  is
    procedure Write (d: Data);
    entry Get (D: out Data);
  private
    Ready: Boolean := false with Synchronized,
      Memory_Order_Read => Acquire,
      Memory_Order_Write => Release;
    Da: Data;
  end RA;

end Generic_Release_Acquire;

package body Generic_Release_Acquire is

  concurrent body RA is

    procedure Write (D: Data) is
    begin
      Da := D;
      Ready := true;
    end Write:

    entry Get (D: out Data)
      when Ready is
      -- spin-lock until released, i.e., Ready = true;
      -- only sync. variables and constants allowed in guard expression
    begin
      D := Da;
    end Get;
  end RA;

end Generic_Release_Acquire;
\end{lstlisting}
\vspace*{-2\baselineskip}
\end{subsubsection}

\end{section}

\begin{section}{API}\label{API_sec}
As already pointed out, we feel that 
providing concurrent objects as first-class citizens is the right way to
enhance Ada with non-blocking synchronization on an adequate memory
model.  On the other hand, if the programmer needs synchronization on a
lower level than \concurrent\ objects provide, an API-based approach
(generic function {\tt Read\_Modify\_Write} in package {\tt
Memory\_Model}) would be a viable alternative.  Listing~\ref{packmm}
shows such a predefined package {\tt Memory\_Model}.  It contains the
specification of generic function {\tt Read\_Modify\_Write}, which
allows to use the read-modify-write operation of the underlying
computer hardware\footnote{An example for employing function {\tt
Read\_Modify\_Write} is given in the Appendix in Listing~\ref{lfsapi}.
It shows an implementation of a lock free stack using generic function
{\tt Read\_Modify\_Write} of package {\tt Memory\_Model}.}.

Exposing sync loops to the programmer makes it necessary to introduce a new aspect {\tt sync\_loop} to let the runtime perform the state change to ``in\_sync\_loop'' (cf.\ Section~\ref{sched}).
Because nobody can force the programmer to use this aspect correctly,
the information transferred to the runtime may be false or incomplete, giving rise to concurrency defects such as
deadlocks, livelocks, and other problems.
\begin{lstlisting}[escapechar=|,
caption={Package Memory\_Model},
label={packmm}
]
package Memory_Model is

  type Memory_Order_Type is (
    Sequentially_Consistent,
    Relaxed,
    Acquire,
    Release);

  subtype Memory_Order_Success_Type is Memory_Order_Type;

  subtype Memory_Order_Failure_Type is Memory_Order_Type
    range Sequentially_Consistent .. Acquire;

  generic
     type Some_Synchronized_Type is private;
     with function Update return Some_Synchronized_Type;
     Read_Modify_Write_Variable: in out Some_Synchronized_Type
       with Read_Modify_Write;
     Memory_Order_Success: Memory_Order_Success_Type :=
       Sequentially_Consistent;
     Memory_Order_Failure: Memory_Order_Failure_Type :=
       Sequentially_Consistent;
  function Read_Modify_Write return Boolean;

end Memory_Model;
\end{lstlisting}
\enlargethispage{2\baselineskip}
\end{section}

\begin{section}{Conclusion and Future Work\label{sec:concl}}

We have presented an approach for providing safe non-blocking synchronization
in Ada~202x.  Our novel approach is based on introducing concurrent objects for
encapsulating non-blocking data structures on a high abstraction level.  In
addition, we have presented \synchronized\ and read-modify-write types which
support the expression of memory ordering operations at a sufficient level of
detail.  Concurrent objects provide SC for programs without data races. This
SC-for-DRF memory model is well-aligned with Ada's semantics for blocking
synchronization via protected objects, which requires legal programs to be
without synchronization errors (\cite[9.10§11]{Ada2012LRM}).

Although Ada~2012 provides the highly abstract protected object
monitor-construct for blocking synchronization, there was previously no
programming primitive available to match this abstraction level for
non-blocking synchronization.  The proposed memory model in conjunction with
our concurrent object construct for non-blocking synchronization may bar users
from having to invent ad-hoc synchronization solutions, which have been found
error-prone with blocking synchronization already~\cite{AdHocSync}. 

Until now, all previous approaches are based on APIs.
We have listed a number of advantages that support our approach of making non-blocking data structures first class language citizens. %
%
In contrast, our approach for Ada~202x encapsulates non-blocking synchronization inside \concurrent\ objects.
This safe approach makes the code easy to understand.
Note that \concurrent\ objects are not orthogonal to objects in the sense of OOP (tagged types in Ada).
However, this can be achieved by employing the proposed API approach (cf.~Section~\ref{API_sec}).
In addition, it is not difficult to migrate code from blocking to non-blocking synchronization.
Adding memory management via storage pools integrates well with our modular approach and does not clutter the code.

A lot of work remains to be done.
To name only a few issues:
Non-blocking barriers (in the sense of~\cite[D.10.1]{Ada2012LRM}) would be useful; details have to be elaborated.
Fully integrating concurrent objects into scheduling and dispatching models and
integrating with the features for parallel programming planned for Ada~202x have to be done carefully.

\end{section}

\section{Acknowledgments}
This research was supported by the Austrian Science Fund~(FWF) project
I~1035N23, and by the Next-Generation Information Computing
Development Program through the National Research Foundation of
Korea~(NRF), funded by the Ministry of Science, ICT \& Future Planning
under grant NRF\-2015\-M3C4A\-7065522.

\bibliographystyle{abbrv} 

\appendix

\begin{section}{Rationale and comparison with C++11\label{sec:rationale}}

We state the rationale for our proposed language features and
compare them to the C++ memory model.
This section thus requires modest familiarity with the C++11 standard~\cite{c++11}.

\subsection{C++11's {\tt compare\_exchange\_weak} and {\tt compare\_exchange\_strong}}
We felt that {\tt compare\_exchange\_weak} and {\tt compare\_exchange\_strong} are not needed on language level.
These are hardware-related details which the compiler knows and should anticipate without intervention of the programmer.

In particular, {\tt compare\_exchange\_weak} means that sometimes a RMW operation fails although the value of the RMW variable has not been changed by a different thread.
In this case re-executing the whole implicit sync loop is not necessary, only the RMW operation has to be redone.
We assume that the compiler produces machine code for this ``inner'' loop.
Because this is only the case on very peculiar CPUs, it is obvious that the compiler and not the programmer should take care of this.

In addition, migrating Ada programs will be facilitated by assigning this job to the compiler.

\subsection{C++11's {\tt consume} memory ordering}
C++ introduced {\tt memory\_order\_consume} specifically for supporting read-copy update (RCU) used in the Linux Kernel (cf.~\cite{preshing:consume}).
However, it turned out that {\tt memory\_order\_consume} as defined in the C++ standard~\cite{c++11}
is not implemented by compilers. Instead, all compilers map it to {\tt memory\_order\_acquire}.
The major reason for this is that the data dependency as defined in~\cite{c++11} is difficult to implement (cf.,~e.g.,~\cite{N4036:14}).
There is, however, ongoing work within ISOCPP to make {\tt memory\_order\_consume} viable for implementation (cf.,~e.g.,~\cite{P0462R1:17,P0190R4:17}).
In particular, ~\cite{P0190R4:17} proposes to restrict {\tt memory\_order\_consume} to data dependence chains starting with pointers
because this represents the major usage scenario in the Linux kernel.

For Ada~202x it seems reasonable not to include {\tt memory\_order\_consume} in the standard.
Instead, compilers are encouraged to exploit features provided by the hardware for gaining performance on weakly-ordered CPUs.
The programmer uses {\tt memory\_order\_release} and {\tt memory\_order\_acquire} for synchronization and the compiler
improves the performance of the program if the hardware is weakly-ordered and it (the compiler) is willing to perform data dependency analysis.
In addition, a compiler switch might be a way for letting the programmer decide whether she is willing to bare the optimization load (increased compile time).

In addition, migrating Ada programs will be facilitated by not having to replace {\tt memory\_order\_acquire} with {\tt memory\_order\_consume} and vice versa depending on the employed hardware.

\subsection{C++11's {\tt acquire\_release} memory ordering}
C++11 defines {\tt acquire\_release} memory ordering because some of C++11's RMW operations contain both a read and a write operation, e.g., {\tt i++} for {\tt i} being an atomic integer variable.
Because Ada's syntax does not contain such operators, {\tt acquire\_release} memory ordering is not needed on language level.
Compiling {\tt i := i+1} ({\tt i} being an atomic integer variable), an Ada compiler is able to employ suitable memory fences to ensure the memory model aspects given by the programmer together with the original statement.

\end{section}
\newpage
\begin{section}{Further Examples}\label{app:examples}

\begin{subsubsection}{Peterson's Algorithm.}
Listing~\ref{peterson} shows an implementation of Peterson's algorithm, a method for lock-free synchronizing two tasks, under the sequentially consistent memory model.

\begin{lstlisting}[escapechar=|,
caption={Peterson's Algorithm under the Sequentially Consistent Memory Model},
label={peterson}
]
concurrent Peterson_Exclusion is
  
  procedure Task1_Critical_Section;
  
  procedure Task2_Critical_Section;

private

  -- Accesses to synchronized variables are atomic and have by default
  -- Sequential Consistency (i.e. the compiler must generate code that
  -- respects program order, and the adequate memory fence instructions
  -- are introduced before and after each load or store to serialize
  -- memory operations in all CPU cores, respecting ordering, visibility,
  -- and atomicity).  However, it is possible to relax each read / write
  -- operation on an synch variable, for obtaining higher performance in
  -- those algorithms that allow these kinds of reorderings by other
  -- threads.

  Flag1 : Boolean := False with Synchronized;
  Flag2 : Boolean := False with Synchronized;
  Turn  : Natural with Synchronized;

  -- Additional data can be placed here, i.e. for shared data variables
  -- that need no atomic accesses (i.e. when data races are not possible
  -- because protected by synchronized variables)

  -- Concurrent entries also encapsulate the access to shared data
  -- (Synch variables), but automatically using spin loops / compare and
  -- swap operations for synchronization among threads.

  entry Task1_Wait;

  entry Task2_Wait;

end Peterson_Exclusion;


concurrent body Peterson_Exclusion is
 
  entry Task1_Busy_Wait
    when Flag2 and then Turn = 2  -- Spin loop until the condition is True
  is
  begin
    null;
  end Task1_Busy_Wait;
 
  entry Task2_Busy_Wait
    when Flag1 and then Turn = 1  -- Spin loop until the condition is True
  is
  begin
    null;
  end Task2_Busy_Wait;

  procedure Task1_Critical_Section is
  begin
    Flag1 := True;
    Turn  := 2;
    Task1_Busy_Wait;
               
    Code_For_Task1_Critical_Section;

    Flag1 := False;
  end Task1_Critical_Section;

  procedure Task2_Critical_Section is
  begin
    Flag2 := True;
    Turn  := 1;
    Task2_Busy_Wait;
       
    Code_For_Task2_Critical_Section;

    Flag2 := False;
  end Task2_Critical_Section;

end Peterson_Exclusion;
\end{lstlisting}

Listing~\ref{peterson_ra_new} shows an implementation of Peterson's algorithm under the release-acquire memory model with default memory model specified in the declarative part.
\begin{lstlisting}[escapechar=|,
caption={Peterson's Algorithm under the Release-Acquire Memory Model with default memory model specified in the declarative part},
label={peterson_ra_new}
]
concurrent Peterson_Exclusion is
  
  procedure Task1_Critical_Section;
  
  procedure Task2_Critical_Section;

private

  Flag1 : Boolean := False
    with Synchronized, Memory_Order_Read => Acquire,
      Memory_Order_Write => Release;
  Flag2 : Boolean := False
    with Synchronized, Memory_Order_Read => Acquire,
      Memory_Order_Write => Release;
  Turn  : Natural
    with Synchronized, Memory_Order_Read => Acquire,
      Memory_Order_Write => Release;

  entry Task1_Wait;

  entry Task2_Wait;

end Peterson_Exclusion;


concurrent body Peterson_Exclusion is
 
  entry Task1_Busy_Wait
    when Flag2 and Turn = 2
    -- Spin loop until the condition is True
  is
  begin
    null;
  end Task1_Busy_Wait;
 
  entry Task2_Busy_Wait
    when Flag1 and Turn = 1
    -- Spin loop until the condition is True
  is
  begin
    null;
  end Task2_Busy_Wait;

  procedure Task1_Critical_Section is
  begin
    Flag1 := True;
    Turn  := 2;
    Task1_Busy_Wait;
               
    Code_For_Task1_Critical_Section;

    Flag1 := False;
  end Task1_Critical_Section;

  procedure Task2_Critical_Section is
  begin
    Flag2 := True;
    Turn  := 1;
    Task2_Busy_Wait;
       
    Code_For_Task2_Critical_Section;

    Flag2 := False;
  end Task2_Critical_Section;

end Peterson_Exclusion;
\end{lstlisting}

Listing~\ref{peterson_ra} shows an implementation of Peterson's algorithm under the release-acquire memory model with memory model explicitly specified at statements.
\begin{lstlisting}[escapechar=|,
caption={Peterson's Algorithm under the Release-Acquire Memory Model with memory model explicitly specified at statements},
label={peterson_ra}
]
concurrent Peterson_Exclusion is
  
  procedure Task1_Critical_Section;
  
  procedure Task2_Critical_Section;

private

  Flag1 : Boolean'Concurrent_Write(Memory_Model => Release) := False
    with Synchronized;
  Flag2 : Boolean'Concurrent_Write(Memory_Model => Release) := False
    with Synchronized;
  Turn  : Natural with Synchronized;

  entry Task1_Wait;

  entry Task2_Wait;

end Peterson_Exclusion;


concurrent body Peterson_Exclusion is
 
  entry Task1_Busy_Wait
    when Flag2'Concurrent_Read(Memory_Model => Acquire) and
      Turn'Concurrent_Read(Memory_Model => Acquire) = 2
    -- Spin loop until the condition is True
  is
  begin
    null;
  end Task1_Busy_Wait;
 
  entry Task2_Busy_Wait
    when Flag1'Concurrent_Read(Memory_Model => Acquire) and
      Turn'Concurrent_Read(Memory_Model => Acquire) = 1
    -- Spin loop until the condition is True
  is
  begin
    null;
  end Task2_Busy_Wait;

  procedure Task1_Critical_Section is
  begin
    Flag1'Concurrent_Write(Memory_Model => Release) := True;
    Turn'Concurrent_Write(Memory_Model => Release)  := 2;
    Task1_Busy_Wait;
               
    Code_For_Task1_Critical_Section;

    Flag1'Concurrent_Write(Memory_Model => Release) := False;
  end Task1_Critical_Section;

  procedure Task2_Critical_Section is
  begin
    Flag2'Concurrent_Write(Memory_Model => Release) := True;
    Turn'Concurrent_Write(Memory_Model => Release)  := 1;
    Task2_Busy_Wait;
       
    Code_For_Task2_Critical_Section;

    Flag2'Concurrent_Write(Memory_Model => Release) := False;
  end Task2_Critical_Section;

end Peterson_Exclusion;
\end{lstlisting}

\end{subsubsection}
\newpage

\begin{subsubsection}{Filter Algorithm.}
\enlargethispage{3\baselineskip}
The filter algorithm is a non-blocking method for synchronizing $n$ processes, which is starvation and deadlock free (\cite{Herlihy:2012}).
Listing~\ref{filter} is an implementation using our proposed approach.
In particular, notice the use of a private entry family.

\begin{lstlisting}[escapechar=|]
generic
  No_Of_Processes: Positive; -- positive number >= 2
package Filter_Algorithm is

  subtype Process_ID is Natural range 0 .. No_Of_Processes-1;
  subtype Process_ID_With_Minus_One is Integer range
    -1 .. No_Of_Processes-1;
  subtype Process_ID_Small is Process_ID range
    Process_ID'FIRST .. Process_ID'LAST-1;
  type Level_Type is array(Integer range <>) of
    Process_ID_With_Minus_One;

  concurrent Access_To_Critical_Section
  is
    procedure Acquire_Lock (ID: Process_ID);
    procedure Release_Lock (ID: Process_ID);
  private
    entry Private_Lock(Process_ID); -- entry family

    Level: Level_Type (Process_ID) := (others => -1)
      with Synchronized_Components,
      Memory_Order_Read => Acquire,
      Memory_Order_Write => Release;
    Last_To_Enter: Level_Type(Process_ID_Small) := (others => -1)
      with Synchronized_Components,
      Memory_Order_Read => Acquire,
      Memory_Order_Write => Release;
    Var_L: Level_Type (Process_ID);

  end Access_To_Critical_Section;

end Filter_Algorithm;
\end{lstlisting}
\begin{lstlisting}[escapechar=|,
caption={Filter Algorithm},
label={filter}
]
package body Filter_Algorithm is

  concurrent body Access_To_Critical_Section is

    procedure Acquire_Lock (ID: Process_ID) is
    begin
      for L in Process_ID_Small'RANGE loop
        Level(ID) := L;
        Last_To_Enter(L) := ID;
        Var_L(ID) := L;
        Private_Lock(ID);
      end loop;
    end Acquire_Lock;

    entry Private_Lock(for ID in Process_ID)
      when ((Last_To_Enter(Var_L(ID)) /= ID) or else
        (for all K in Level'RANGE => ( K /= ID and then
          Level(K) < Var_L(ID))))
    is
    begin
      null;
    end Private_Lock;

    procedure Release_Lock (ID: Process_ID) is
    begin
      Level(ID) := -1;
    end Release_Lock;

  end Access_To_Critical_Section;

end Filter_Algorithm;
\end{lstlisting}

\end{subsubsection}

\newpage
\begin{subsubsection}{API-based non-blocking stack.} Here we present how a non-blocking stack can be implemented via the API proposed in Sec.~\ref{API_sec}.

\begin{lstlisting}[escapechar=|,
caption={Non-blocking Stack Implementation Using Generic Function\newline {\tt Memory\_Model.Read\_Modify\_Write}},
label={lfsapi}
]
  subtype Data is Integer;

  type List;
  type List_P is access List;
  type List is
    record
      D: Data;
      Next: List_P;
    end record;

  Empty: exception;

  concurrent Lock_Free_Stack
  is
    procedure Push(D: Data);
    procedure Pop(D: out Data);
  private
    Head: List_P with Read_Modify_Write;
  end Lock_Free_Stack;

  concurrent body Lock_Free_Stack is
    procedure Push (D: Data) is
      New_Node: List_P := new List;
      function Update_Head_Push return List_P is
      begin
        return New_Node;
      end Update_Head_Push;
      function RMW_Head_Push return Boolean is
        new Memory_Model.Read_Modify_Write(
          Some_Synchronized_Type => List_P,
          Update => Update_Head_Push,
          Read_Modify_Write_Variable => Head,
          Memory_Order_Success => Release,
          Memory_Order_Failure => Relaxed);
    begin
      loop with Sync_Loop
        New_Node.all := (D => D, Next => Head'Concurrent_Read(
          Memory_Order => Relaxed);
        exit when RMW_Head_Push;
        -- This is an RMW operation; so: if value of head has changed in
        -- between, the loop is reexecuted;
        -- if not, the assignment succeeds.
        -- NOTE: memory_order release initiates a happens_before
        -- relationship for the memory_order aquire in pop
      end loop;
    end Push;

    procedure Pop(D: out Data) is
      Old_Head: List_P;
      function Update_Head_Pop return List_P is
      begin
        return Old_Head.Next;
      end Update_Head_Pop;
      function RMW_Head_Pop return Boolean is
        new Memory_Model.Read_Modify_Write(
          Some_Atomic_Type => List_P,
          Update => Update_Head_Pop,
          Read_Modify_Write_Variable => Head,
          Memory_Order_Success => Relaxed,
          Memory_Order_Failure => Relaxed);
    begin
      loop with Sync_Loop
        Old_Head := Head'Concurrent_Read(Memory_Order => Relaxed);
        if Old_Head /= null then
          if RMW_Head_Pop then
          -- This is an RMW operation; so: if value of head has changed in
          -- between, the if statement terminates and the loop body is
          -- executed once more,
          -- if not, the assignment succeeds and the then branch is
          -- executed.
          -- NOTE: memory_order aquire establishes a happens_before
          -- relationship with the memory_order release in push
            D := Old_Head.D;
            exit;
          end if;
        else
          raise Empty;
        end if;
      end loop;
    end Pop;
  end Lock_Free_Stack;
\end{lstlisting}

\end{subsubsection}

\end{section}
\end{document}